\documentclass[12pt,preprint]{aastex}

\newcommand{\ROSAT}{{\it ROSAT\ }}
\newcommand{\OVI}{O\ {\footnotesize VI}}
\newcommand{\OVII}{O\ {\footnotesize VII}}

\shorttitle{}
\shortauthors{Mittaz et al.}

\begin{document}


\title{WHIM emission and the cluster soft excess: a model comparison}


\author{J. Mittaz, R.Lieu}
\affil{Department of Physics, University of Alabama at Huntsville, Huntsville, AL 35899}


\author{R. Cen}
\affil{Peyton Hall, Princeton University, Princeton, NJ 08544}

\and

\author{M. Bonamente}
\affil{Department of Physics, University of Alabama at Huntsville, Huntsville, AL 35899}




\begin{abstract}
The confirmation of the cluster soft excess (CSE) by XMM-Newton has rekindled
interest as to its origin.  The recent detections of CSE emission at large
cluster radii together with reports of \OVII\ line emission associated with the
CSE has led many authors to conjecture that the CSE is, in fact, a signature of
the warm-hot intergalactic medium (WHIM).  In this paper we test the scenario
by comparing the observed properties of the CSE with predictions based on
models of the WHIM.  We find that emission from the WHIM in current models is 3
to 4 orders of magnitude too faint to explain the CSE.  We discuss different
possibilities for this discrepancy including issues of simulation resolution
and scale, and the role of small density enhancements or galaxy groups.  Our
final conclusion is that the WHIM alone is unlikely to be able to accout for
the observed flux of the CSE.
\end{abstract}




\section{Introduction}
The location of all the baryons existing at the current epoch is still somewhat
of a mystery.  Observationally, the total sum of baryons seen in stars,
galaxies and clusters of galaxies ($\Omega_b = (2.1^{+2.0}_{-1.4})
h_{70}^{-2}$\%, Fukugita, Hogan \& Peebles 1998) is only about half of the
number of baryons required by big bang nucleosynthesis models ($\Omega_b = (3.9
\pm 0.5) h_{70}^{-2}$\% Burles \& Tytler 1998) or from measurements of the
cosmic microwave background ($\Omega_b = (4.4 \pm 0.4) h_{70}^{-2}$\% Bennett
et al. 2003).  Recent cosmological hydrodynamical simulations have, however,
shown that this missing 50\% of baryons may be in the form of a warm ($10^5 -
10^7$K), tenuous medium (with overdensities between $\delta \sim 5-50$)
existing in filaments formed during the process of large scale structure
formation (e.g. Cen \& Ostriker 1999).  This medium is generally called the
'warm-hot intergalactic medium' or WHIM.

Ever since it was first proposed, the detection of the WHIM has been an
important goal in astrophysics.  To date there has been a considerable amount
of success in finding WHIM like material, particularly from absorption line
studies.  Far UV and soft X-ray absorption lines from the WHIM have been
reported for both the local group as well as at higher redshifts e.g. Fang,
Sembach \& Canizares 2003; Nicastro et al. 2003; Mathur et al. 2003, Tripp,
Savage \& Jenkins 2002.  Possible emission from the the WHIM has also been
observed, with some weak X-ray detections (e.g. Soltan, Freyberg \& Hasinger
2002; Zappacosta et al. 2002) which are in approximate agreement with the
predicted luminosity of the WHIM inferred from cosmological simulations.

Recently, new observations of the cluster soft excess emission (CSE) show that
this phenomenon may also be a signature of the WHIM.  The CSE (seen as an
excess of observed flux above the the hot intracluster medium (ICM) at energies
below lkeV) has been an observational puzzle for a number of years.  First
discovered in the Virgo cluster (Lieu et al. 1996) subsequent observations have
found similar behavior in a number of different systems (e.g. Lieu et al.
1996b; Mittaz et al. 1998; Kaastra et al. 1999; Bonamente et al. 2001a;
Bonamente et al.  2001b; Bonamente et al. 2002).  Observationally the CSE shows
a number of different characteristics.  In some clusters the CSE when expressed
as a percentage is spatially constant (e.g. Coma: Lieu et al. 1998), in others
there is a marked radial dependence with the fractional soft excess being
stronger in the outer regions of the cluster (e.g. Mittaz et
al. 1998). Observations have also shown that the CSE is a relatively common
phenomena - a \ROSAT study of 38 clusters showed that approximately 45\% of
clusters show at least a $1\sigma$ effect (Bonamente et al. 2002) .

On the current state of interpretation, models assuming a thermal origin of the
excess have invoked a warm gas intermixed with the hot ICM are unsatisfactory.
These have found to be unsatisfactory, however, since the cooling times are
extremely short (sometimes $\sim 10^6$ years Mittaz et al. 1998) so some kind
of heating mechanism to sustain the warm gas has to be envisaged (Fabian 1997).
On the other hand if the emission arises from an inverse-Compton process (such
as suggested by Hwang 1997; Sarazin \& Lieu 1998; Ensslin, Lieu \& Biermann
1998) then the radial dependence seen in some clusters can be explained, but
the inferred pressure of the required cosmic-rays would be perplexedly too high
(Lieu, Axford \& Ip 1999).

The picture has changed somewhat with new observations by the XMM-Newton
satellite.  These observations reveal apparent strong thermal lines of Oxygen
in the CSE spectra for the outskirts of a number of clusters (Kaastra et
al. 2003; Finoguenov et al. 2003) with the CSE continuum being fitted with a
characteristic temperature of 0.2 keV.  If established, this provides
irrefutable evidence that the CSE must be thermal in nature.  Such a finding is
also supported by \ROSAT observations of the Coma cluster, which shows a very
large, degree scale halo of soft emission (Bonamente et al. 2003).  Emission on
this scale cannot possibly be non-thermal in nature since there is no way of
confining a relativistic particle population at such distances from the cluster
center.  Therefore, given that there seems to be a large scale thermal
component at the outskirts of clusters, the CSE emitting material has to be
located predominantly beyond the cluster virial radius where we can have
sufficiently low densities to overcome any cooling time issues.  In such a
scenario the warm emitting gas is spatially consistent with the WHIM
i.e. current cosmological models should be able to predict the luminosity and
temperature of the CSE.  We have therefore undertaken a detailed comparison
between cosmological simulations of the WHIM and the observed soft excess
signal.

\section{The model}

We have used a recent cosmological hydrodynamic simulation of the canonical
cosmological constant dominated cold dark matter model (Ostriker \& Steinhardt
1995) with the following parameters: $\Omega_m$ = 0.3, $\Omega_\Lambda$ = 0.7,
$\Omega_b$ $h^2$ = 0.017, h = 0.67, $\sigma_8 = 0.9$ and a spectral index of
the primordial mass power spectrum of n = 1.0. The simulation box has a size of
25 h$^{-1}$ Mpc comoving on a uniform mesh with 768$^3$ cells and 384$^3$ dark
matter particles giving a comoving cell size of 32.6 h$^{-1}$ kpc. This
simulation together with another similar one at lower resolution derived from
the same code and parameters have previously been used to account for a variety
of of different observational consequences of the WHIM, such as \OVI\
absorption line studies (e.g. Cen, Tripp, Ostriker, Jenkins, 2001) and the
X-ray background (Phillips, Ostriker \& Cen, 2001).  For a more detailed
discussion of the simulation itself see Cen, Tripp, Ostriker \& Jenkins (2001).
Here we are primarily concerned with the emission in the vicinity of a cluster.
Thus we focus our analysis on a cluster simulated in the high resolution mode.

\clearpage

\begin{figure}
\centerline{\includegraphics[height=3.0in,angle=0]{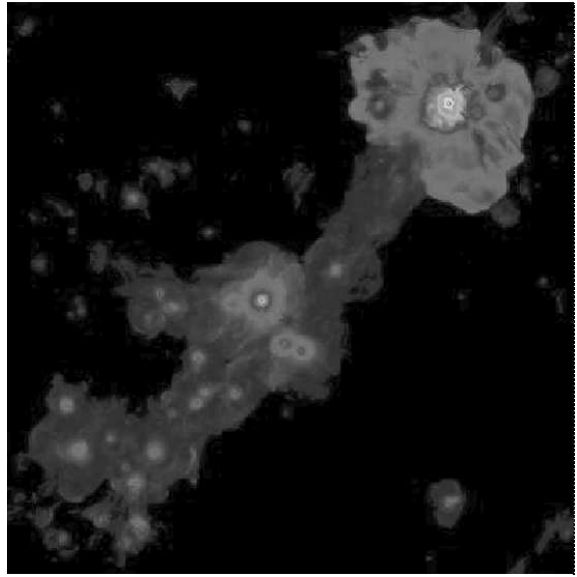}}
\caption{Emission weighted temperature map for one particular projection of the
  data volume.  The plotted temperature range is from 0.1-6.5 keV.  The
  simulated cluster can be seen in the top right of the figure}\label{fig1}
\end{figure}

\clearpage

The simulation as provided comes in the form of three data cubes containing
temperature, density and metal abundance.  Figure~\ref{fig1} shows an emission
weighted temperature map from one particular projection of the cube and within
the image a number of structures can be seen, including filaments, groups and
one cluster candidate at a temperature of $\sim$ 6.5keV.  For the purpose of
this paper we are going to concentrate on this cluster-like structure which is
seen in the top right corner of Figure~\ref{fig1}.  It has a peak mass density
of $5.7 \times 10^{13}$ M$_\odot$ Mpc$^{-2}$ placing it the regime of
virialized objects and so can be considered for our present purposes a cluster
of galaxies.
The parameters of this simulated cluster are those of a relatively massive
cluster and such clusters do show soft excesses.  In particular Coma (a very
massive cluster) showed one of the first CSEs, and CSEs are found in clusters
with a wide range of temperatures and masses.  Therefore, there is no reason to
suppose that this structure is inappropriate for looking for the signature of a
WHIM produced soft excess.

\section{Simulated X-ray spectra from the model}

\clearpage

\begin{figure}[t]
\begin{center}
\begin{minipage}[b]{5in}
\centering
\includegraphics[height=3in,angle=-90]{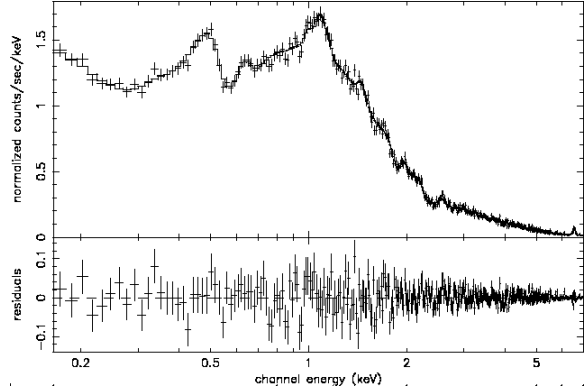}
\end{minipage}
\end{center}
\caption{The simulated spectrum from the cluster showing the spectrum from the
central 32.6 kpc region (0-2 arcminutes).  Also shown is the best fit model (kT
= $4.7 \pm 0.05$ Abundance=0.5) together with fit residuals and
shows no requirement for any cluster soft excess}\label{fig2a}
\end{figure}

\begin{figure}[t]
\begin{center}
\begin{minipage}[b]{5in}
\includegraphics[height=2.2in,angle=-90]{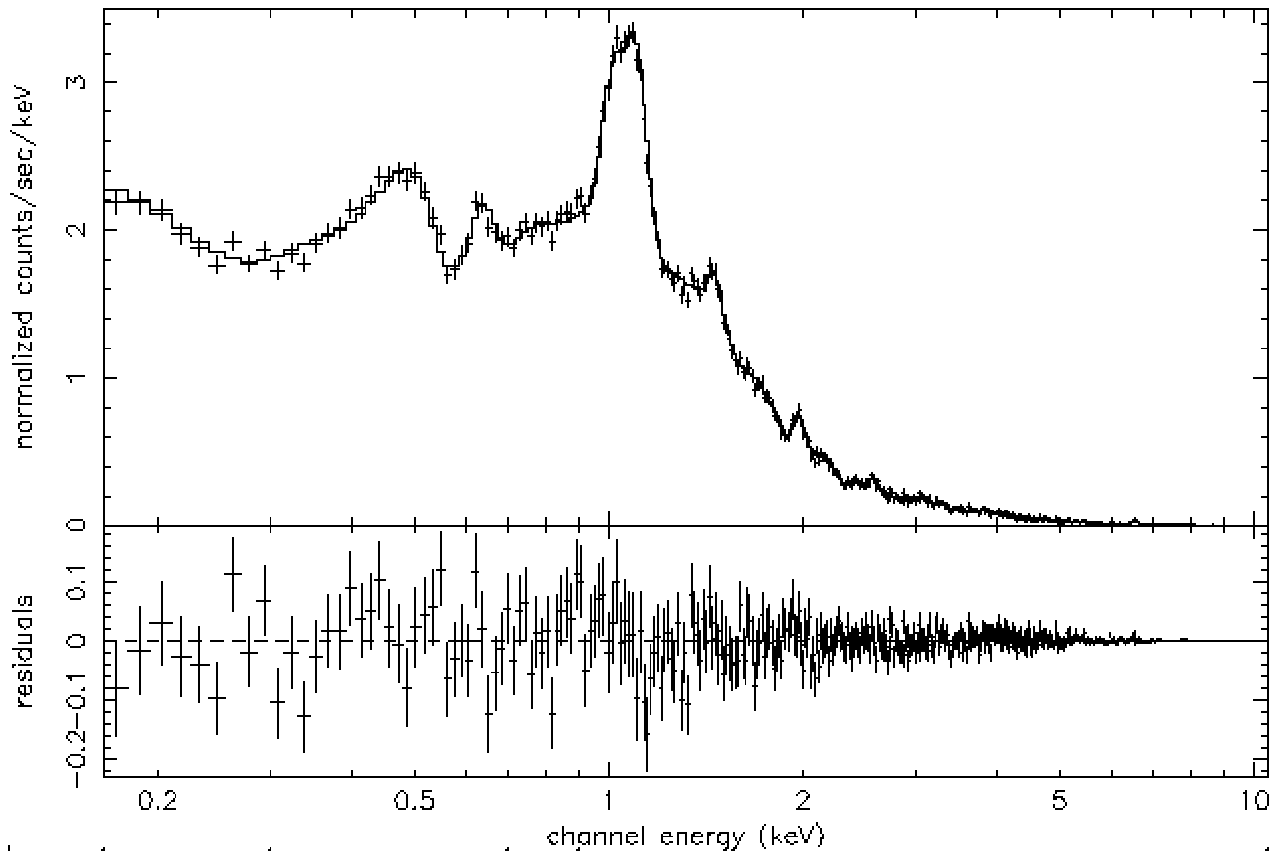}
\hspace{0.2cm}
\includegraphics[height=2.2in,angle=-90]{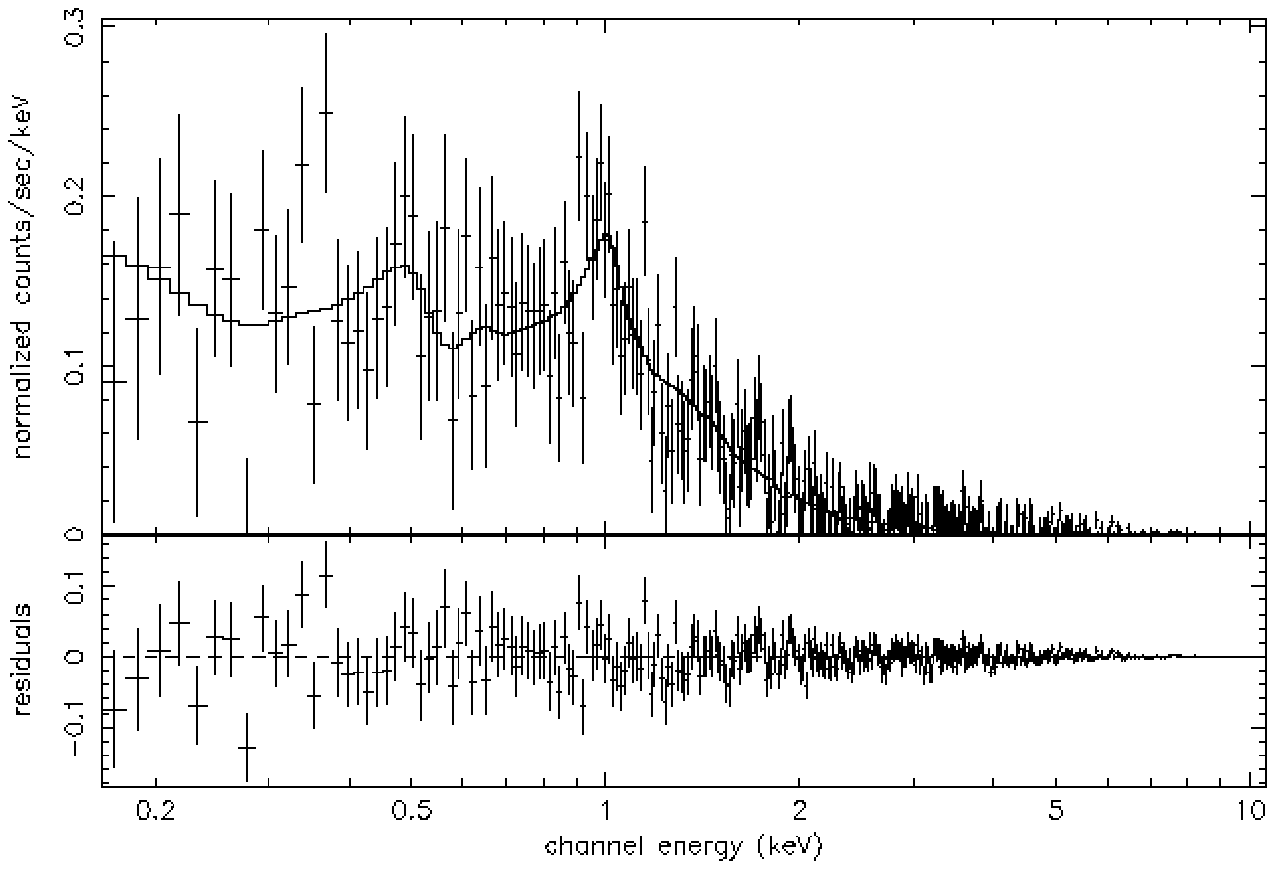}
\end{minipage}
\end{center}
\caption{Further spectra from the simulated cluster from annular
regions placed at the outskirts of the cluster.  The left panel shows
the 589-680 kpc annulus, and the right panel shows the simulated
spectrum taken from the 1.17-1.26 Mpc annulus.  Again, all spectra are
well fitted with a single temperature model (kT = $2.27 \pm 0.08$ keV
Abundance=0.29
and kT=$1.25 \pm 0.13$ Abundance=0.06 respectively) with no evidence for any soft excess
emission.}\label{fig2}
\end{figure}

\clearpage

The ultimate aim of this paper is to investigate the WHIM model predicted X-ray
emitting properties and compare them with current observations, particularly
those of the CSE as now observed with XMM-Newton.  To do this we have placed
the simulation at a redshift of 0.02, similar to that of the Coma cluster whose
CSE has been extensively studied (Bonamente et al. 2003; Finoguenov et al.
2003).  We further assumed that the gas is in thermal and ionization
equilibrium.  Since the simulation provides the temperature, density and metal
abundance at each point, for each cell we can derive an optically thin X-ray
spectrum.  To do this we have used the mekal code for optically thin plasmas
(Kaastra et al. 1992, Leidhal et al. 1995) to generate the average
bremsstrahlung and emission-line spectra.  In order to avoid possible extreme
abundance values that can sometimes be found in the densest regions of the
simulation we do not allow the metal abundance to exceed a value of 0.5 solar.
Then after adding the appropriate galactic absorption (N$_H = 9 \times 10^{19}$
cm$^{-2}$) and folding the resultant spectrum through the XMM-Newton MOS1
response we can generate an XMM-Newton counts spectrum from each cell.  To
obtain the overall spectrum we simply sum these spectra along our given line of
sight.  
Finally, in order to get a reasonable estimate of the noise we must taken into
account the effect of background subtraction on the determination of the CSE.
To do this we have determined what the astrophysical background spectrum is by
using the best fit model parameters for the cosmic X-ray background taken from
Lumb et al. (2002).  The background model includes both Galactic and
extra-galactic backgrounds and the normalisations were scaled by the extraction
area for each spectrum.  Poisson noise was then added to the total spectrum
assuming an exposure time of 50 ksec exposure.  Then a simulated background
spectrum (again with appropriate Poisson noise) was subtracsted from the total
spectrum in the same way as would be done with a real observation.

\clearpage

\begin{figure}[t]
\centering
\includegraphics[height=2.2in,angle=-90]{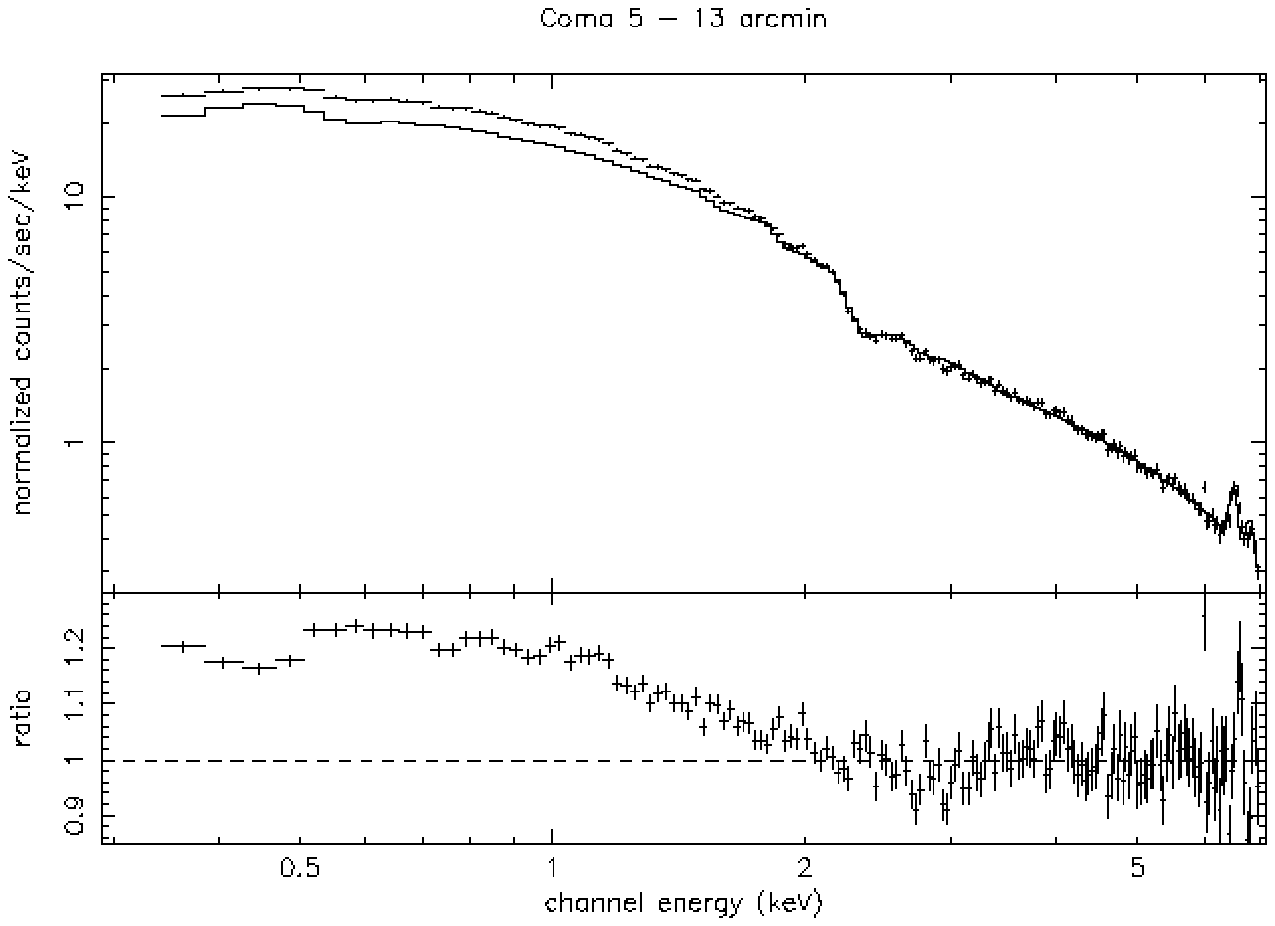}
\hspace{0.2cm}
\includegraphics[height=2.2in,angle=-90]{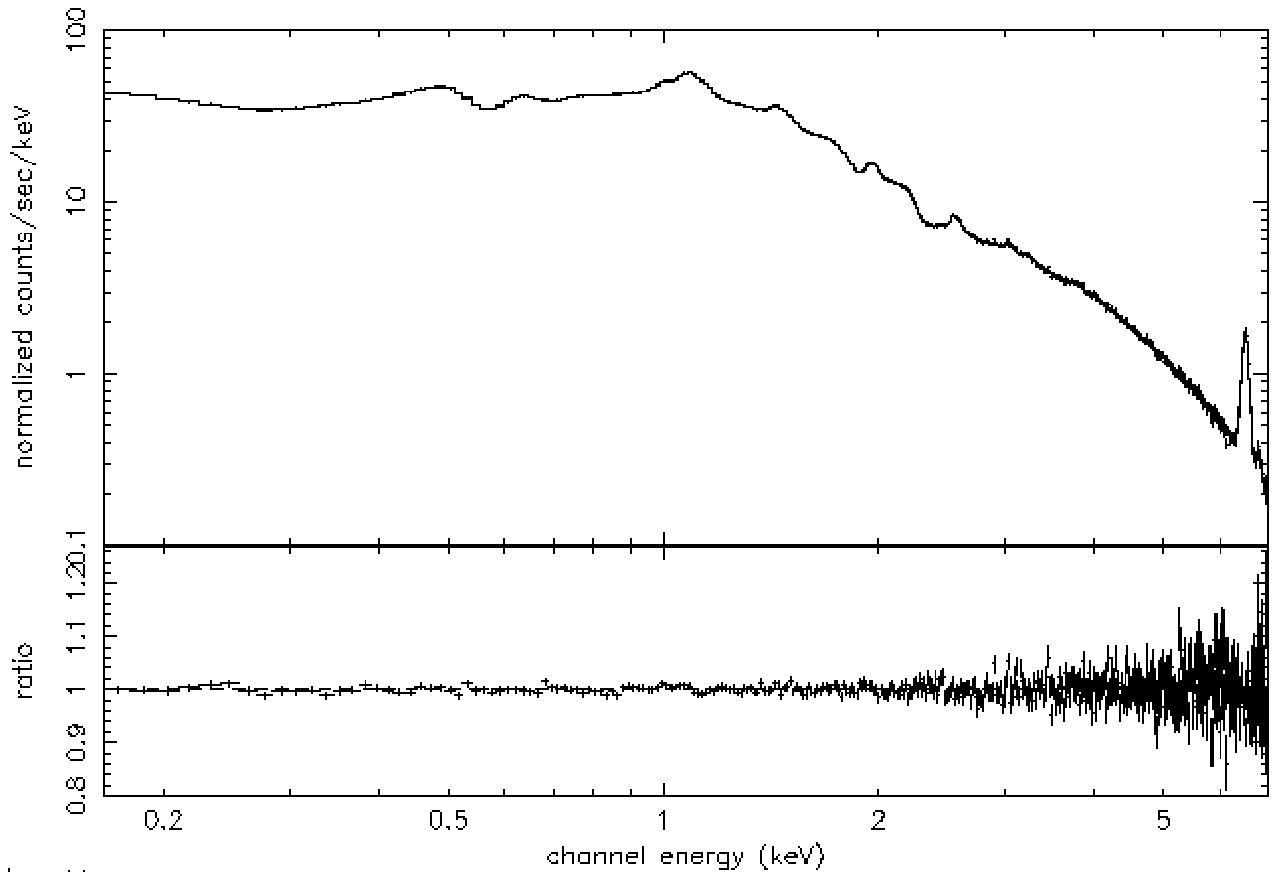}
\caption{The left panel shows the XMM-Newton spectrum of Coma extracted in the
  5-13 arcminute region (from Nevalainen et al. 2003), and the right panel
  shows a simulated spectrum from the model extracted from a similarly sized
  region.  While the observed data shows a clear soft excess, the spectrum from
  the simulation show no need for any extra components above the hot
  ICM emission.}\label{fig5}
\end{figure}

\clearpage

Figure~\ref{fig2a} shows an example of a simulation of the central 40 kpc
(i.e. one cell which corresponds to 0-2 arcminutes) region of the cluster.
Also shown is the best fit performance of the single temperature mekal emission
model to the data with parameters of $4.7 \pm 0.02$ keV and abundance of 0.5.
As can be seen from the residual plot, there are no strong deviations of the
data from the fit. This is consistent with general properties of clusters
inferred from XMM-Newton and Chandra data (e.g. Molendi \& Pizzolato 2001)
where observationally there is no requirement for a multi-temperature fit, even
within the so called cooling radius of the cluster where there is presumed to
be a wide range of temperatures along the line of sight.  As pointed out by
Molendi \& Pizzolato (2001) and Ettori (2002) this is to be expected for a
single phase medium which further supports our model choice.

For a more in depth understanding, the reason why we see such a good fit to the
data even though there are multiple temperatures and densities along the line
of sight, is due to the weighting effect of density.  Since emission goes as
$n_e^2$, the spectrum is heavily biased towards the densest material along the
line of sight.  Therefore the spectrum is dominated by the emission from a
narrow range of densities and temperatures.  Other components, such as may give
rise to a CSE, will not be visible unless they exists in substantial quantities
relating to the principle component.

\subsection{Looking for the cluster soft excess}

The sufficiency of a single temperature fit to the X-ray spectrum is seen at
almost all radii ranging from the center to the outermost regions. This is
illustrated in Figure~\ref{fig2} which shows examples from two outer annuli
situated at 580-680 kpc and 1.17-1.26 Mpc from the center where the one
temperature fits (with kT = $2.27 \pm 0.08$ keV \& kT = $1.25 \pm 0.13$ keV
respectively) are perfectly adequate.  In the context of a WHIM explanation of
the CSE this is interesting, because if the cluster soft excess is due to
overlying filaments from the WHIM it is exactly in these outer regions (where
the cluster emission is weak) that we would expect to see an effect.  The fact
the we see no evidence for a soft excess implies that the emission from the
filaments in this simulation must be much weaker than that of the observed CSE.

\clearpage

\begin{figure}
\centering
\includegraphics[height=3in,angle=-90]{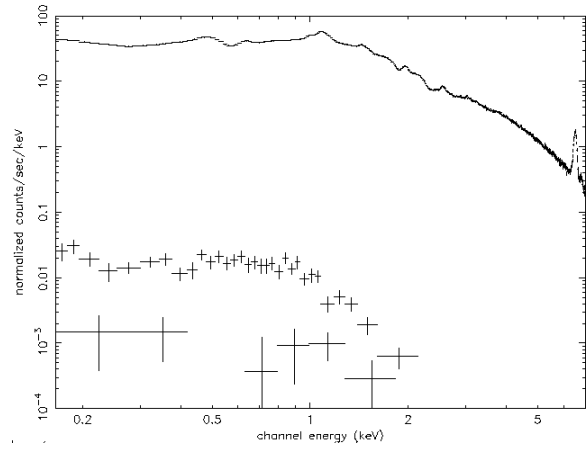}
\caption{The total simulated spectrum from figure~\ref{fig5} (top curve)
  together with spectra derived from cells with temperatures below 1\ keV.  The
  lowest spectrum is that derived from the same line of sight used for the
  total spectrum, while the middle spectrum is taken from a line of sight
  chosen to maximize the soft excess emission.}\label{fig6}
\end{figure}

\clearpage

The difference between observations and theory is even more clearly seen in
figure~\ref{fig5} which shows two spectra: the spectrum on the left is taken
from Nevalainen et al. (2002) and is of the 5-13 arcminute region of the Coma
cluster, the spectrum on the right is taken from the simulation covering a
similar annular region.  The first point to note is the relatively good
agreement in the flux levels - the 0.2 - 2keV luminosity from the simulation is
$\sim 4.5 \times 10^{44}$ ergs/s while the real observations shows a luminosity
of $\sim 8.8 \times 10^{44}$ ergs/s, giving us confidence that the cross
normalizations are realistic.  However, there are also significant differences.
Below 2 keV the XMM-Newton observation of Coma shows a clear soft excess at a
25\% level above the hot ICM emission, whereas the simulated spectrum show no
soft excess emission.  We therefore conclude that the model does not include
material at the right temperature and density to account for the thermal origin
of the CSE.

Such a conclusion is further underpinned when the emissivity of material at a
temperature consistent with the CSE is studied.  Figure~\ref{fig6} shows the
same simulated spectrum as the one in figure~\ref{fig5} but with the addition
of spectra derived only from those cells with temperatures below 1\ keV
i.e. just those cells from which we would expect the cluster soft excess
emission to arise.  The two lower spectra in figure~\ref{fig6} show the
expected emission from all components below 1\ keV as taken from two different
lines of sight.  The lower spectrum is derived from the same sight line used to
create the total spectrum (i.e. the top spectrum).  The brighter of the two
spectra is taken from a sight line that has the most CSE effect, because along
it the emission from cells with a temperature less the 1\ keV is maximized i.e
a line of sight that maximizes the CSE.  Note in both cases there {\it is}
emission in the crucial $< 1$\ keV regime only it is much fainter (by a factors
of $10^4$ or $10^{3}$ respectively) than the hot ICM emission.  It then becomes
obvious why no cluster soft excess was seen in the simulated spectra: the
emission from cells capable of accounting for the soft excess is very small
compared with the soft flux from the hot ICM emission lying behind it.  We can
further quantify this effect by comparing our model prediction with the soft
excess emission seen in a large sample of clusters.  Bonamente et al. (2002)
studied 38 clusters and listed the strength of any soft excess detected with
\ROSAT.  The simulated CSE luminosities fall short of the values for the
majority of these cases.

\subsection{A possible detection of a soft excess and the importance of small
groups}\label{group}

\begin{figure}[t]
\centering
\includegraphics[height=2.2in,angle=-90]{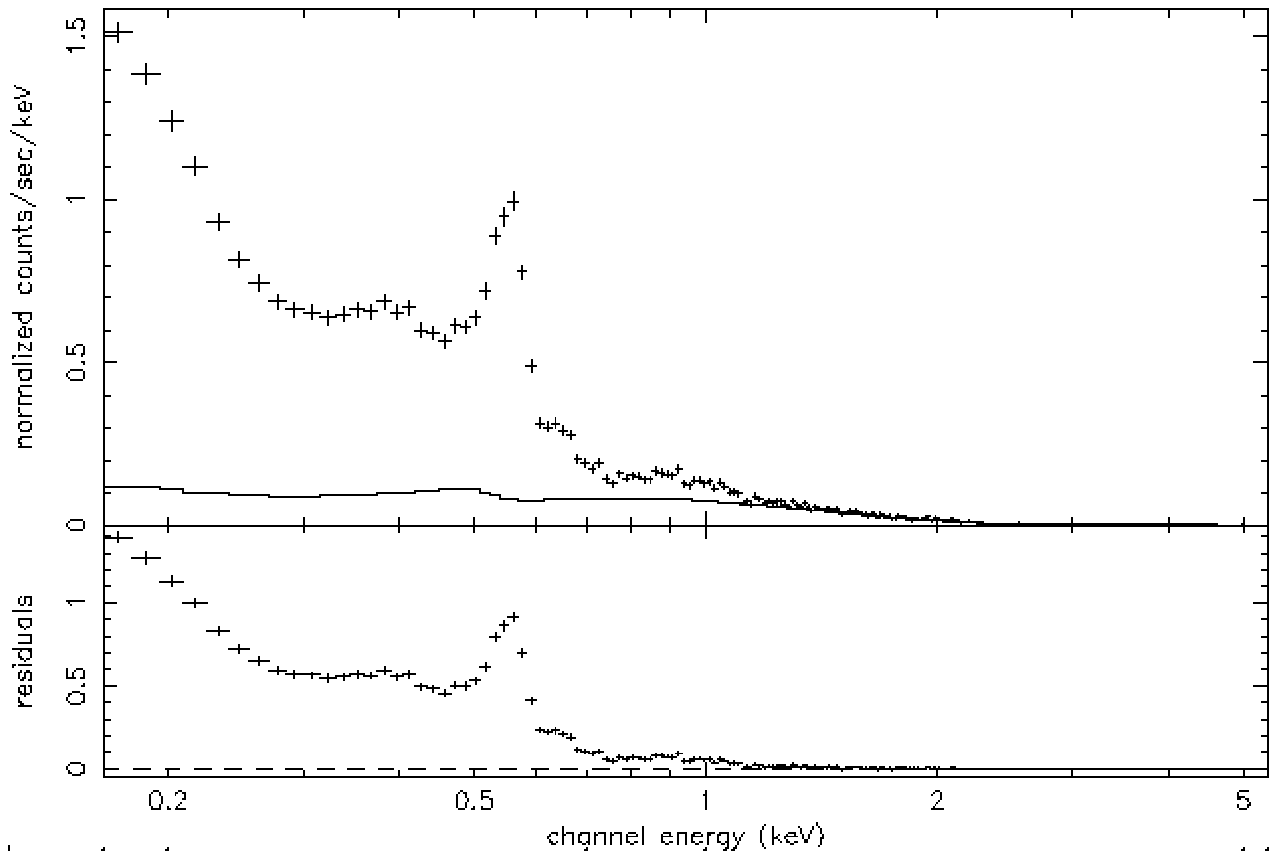}
\hspace{0.2cm}
\includegraphics[height=2.2in,angle=-90]{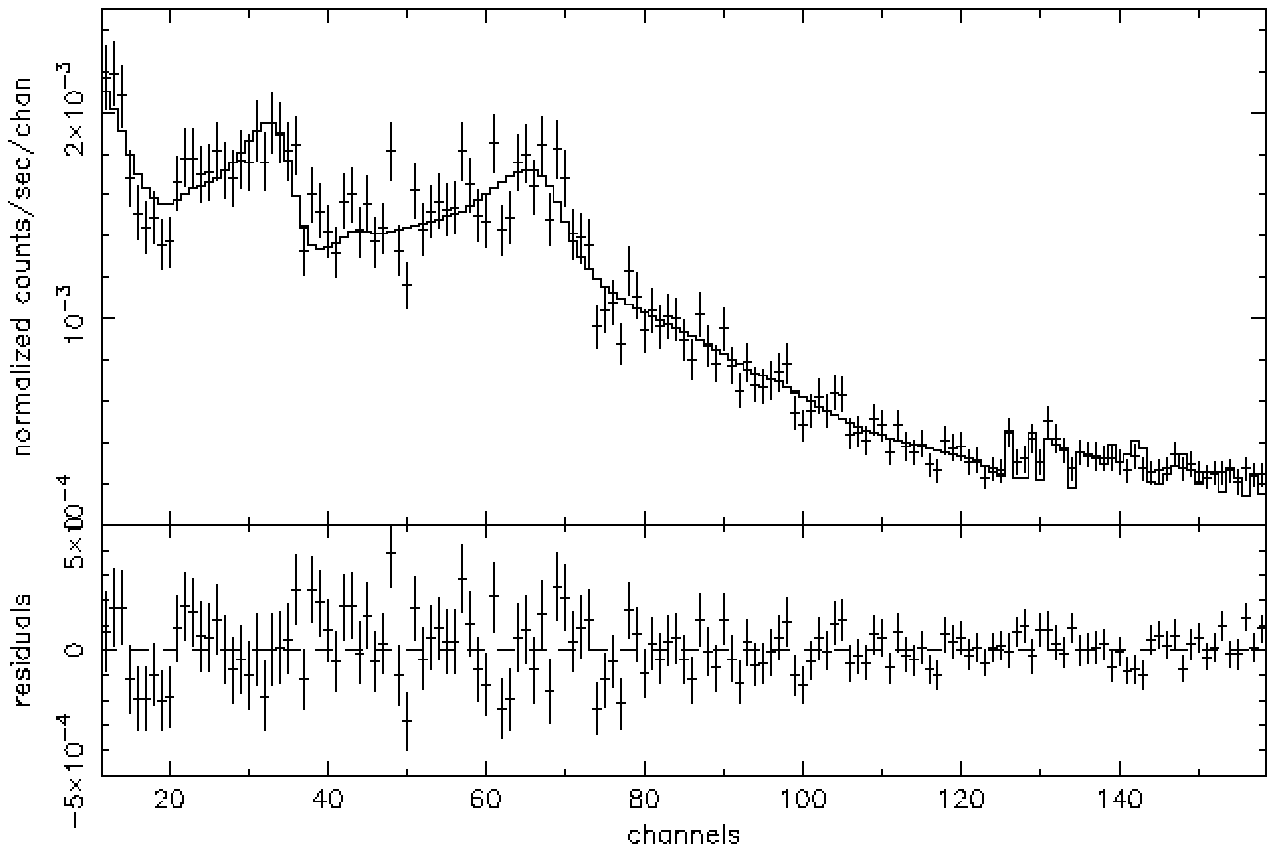}
\caption{The left panel shows the simulated XMM-Newton spectrum for the annular
  region 978-1467 kpc showing a very strong soft excess signal.  The right
  panel shows the same region if a small region containing a strong density
  enhancement is excluded.  The fit now is acceptable over the entire
  XMM-Newton energy range ($\chi^2_{\nu} = 1.18$) with a temperature of 1.07
  keV}\label{fig9}
\end{figure}

\begin{figure}[t]
\centering
\includegraphics[height=2.2in]{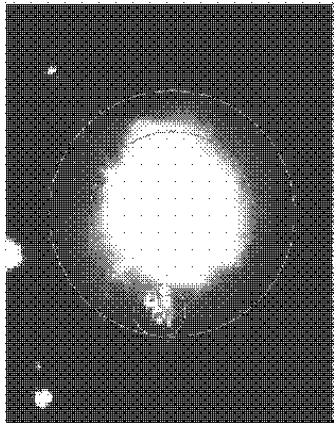}
\caption{An image of the simulated cluster showing the 978-1467 kpc
  extraction annulus used to extract the spectrum from
  figure~\ref{fig9} together with the excluded region that contains
  all of the soft excess emission.  The maximum count rate shown in
  the image is 0.01 cts/sec/pixel.}\label{fig10}
\end{figure}

Under certain circumstances it {\it is} possible to emulate a soft excess.  For
example from an extraction from the 30 - 45 cell (978 - 1467 kpc) annular
region one notices a strong excess at the lowest energies.  The left panel of
Figure~\ref{fig9} shows the simulated spectrum, and in this case a single
temperature fit is unacceptable with a $\chi^2_{\nu}$ of 42.5.  Following the
analysis techniques of Nevalainen {\it et al.}  (2002) and Bonamente {\it et
al.} (2003) of applying a one temperature model to the spectrum above $\sim 1$
keV we find a reasonable fit with a single temperature of $1.08 \pm 0.05$ keV
($\chi^2_{\nu} = 1.28$).  Below 1 keV there is then a very strong soft excess
including a very strong \OVII\ line.  Interestingly the temperature of this
excess is 0.2 keV, exactly the same temperature as has been reported for a
number of clusters (eg. Kaastra {\it et al.} 2003).  However, it turns out that
this excess arises from a very small number of cells within the extraction
region.  In Figure~\ref{fig10} an image of the outskirts of the simulated
cluster together with the annulus used to extract the original spectrum.  Also
shown is a small circle on the left hand side containing the region of bright
pixels i.e. high densities, that actually gives rise to the soft excess.  If
this small region is removed, the resultant spectrum (the right hand panel of
Figure~\ref{fig9}) yields no evidence of a CSE.

We therefore conclude that there is very little evidence to link emission from
the WHIM with the cluster soft excess.  Our investigation has revealed the
possibility, however, that small density enhancements (which may be associated
with small galaxy groups) can give a signal that mimics the cluster soft
excess.  However, in the real observations the cluster soft excess seems
relatively smooth and does not seem to be confined to a few small locations
(see for example the \ROSAT results on the Coma cluster Bonamente {\it et al.}
2003).  One possibility is that the models to date do not have sufficient
spatial resolution to emulate a large population of small density perturbations
which could give rise to the observed soft excess signal.  Additional physics
related to a more complex environment, such as would exist in a supercluster,
may also be required to achieve this - Kaastra et al. (2003) discussed the
possibility, which clearly awaits further work.  An overall fundamental problem
exists, though, and it pertains to the acute lack of mass in gas at any
temperature range to account for the soft excess.  To get the required factor
of $10^3$ increase in brightness from the WHIM would imply an increase in the
density by a factor of 30.  Since the majority of the WHIM has overdensities
between 10-30 we then require overdensities greater than 300.  However, only
$<20$\% of the WHIM bave overdensities greater than 200 (e.g. Dav\'e et
al. 2001) and it is clear that the models simply do not contain material at the
densities and temperatures to give rise to the CSE signal as observed.

\clearpage

\clearpage

\section{The importance of superclustering and the CSE}\label{supercluster}

Another possible explanation for the CSE emission has been raised by Kaastra
{\it et al.}  (2003,2004) who have proposed a correlation between the cluster
soft excess and a supercluster environment.  They arrive at this conclusion
based on two arguments.  \ROSAT all-sky survey maps on degree scales around
soft excess clusters seem to show a large scale excess of soft emission which
is claimed to be related to a supercluster.  The authors also claimed that in
the model of Fang et al. (2002), regions where there are numerous structures
(i.e. a potential simulated supercluster) can also be associated with a \OVII\
column density similar to that inferred from observations ($\sim 4 - 9 \times
10^{16}$ cm$^{-2}$) thereby implying a causal link between superclusters and
the presence of a CSE.

\begin{figure}[t]
\centering
\includegraphics[height=2.5in,angle=-90]{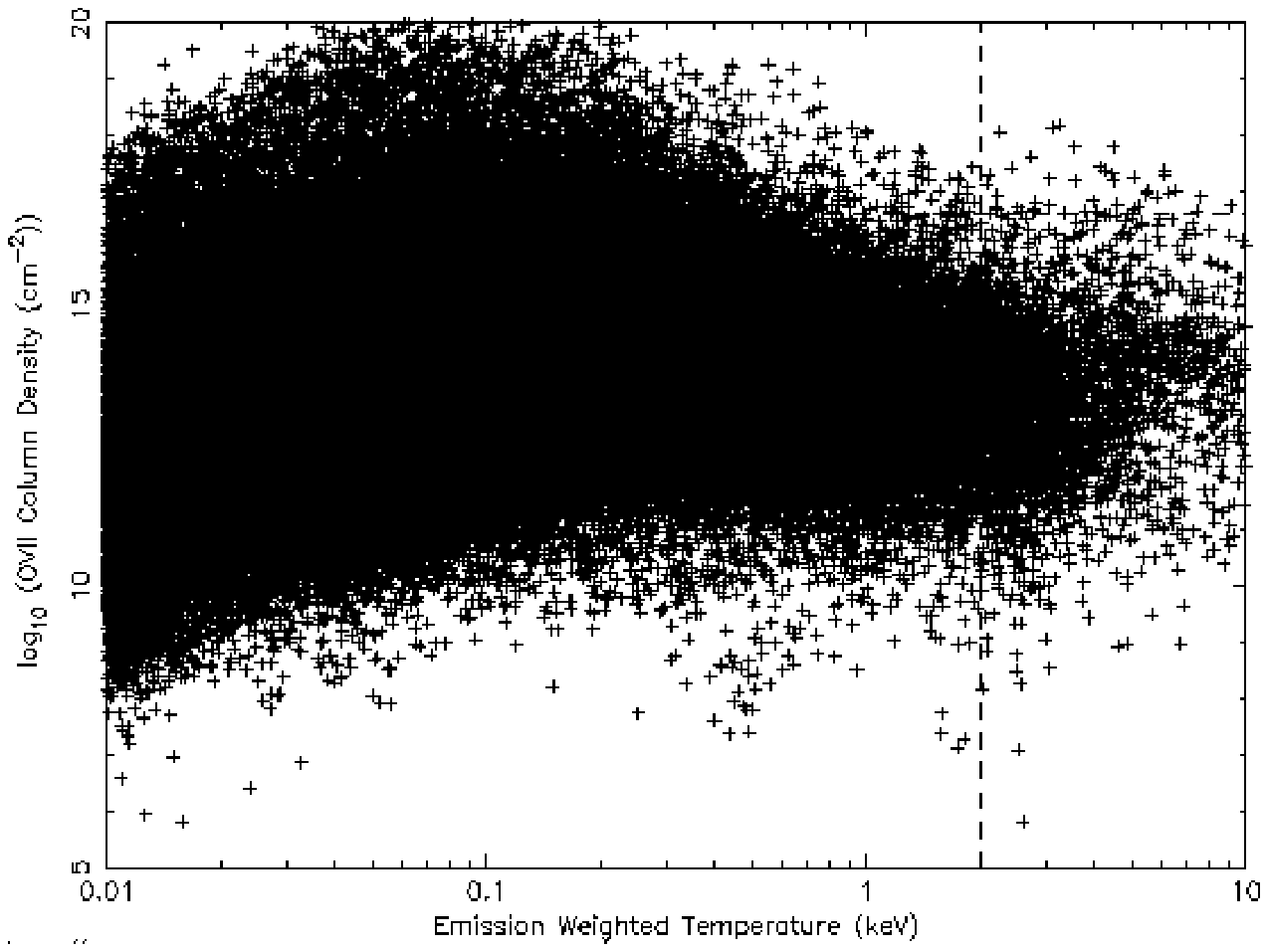}
\hspace{0.2cm}
\includegraphics[height=2.5in,angle=-90]{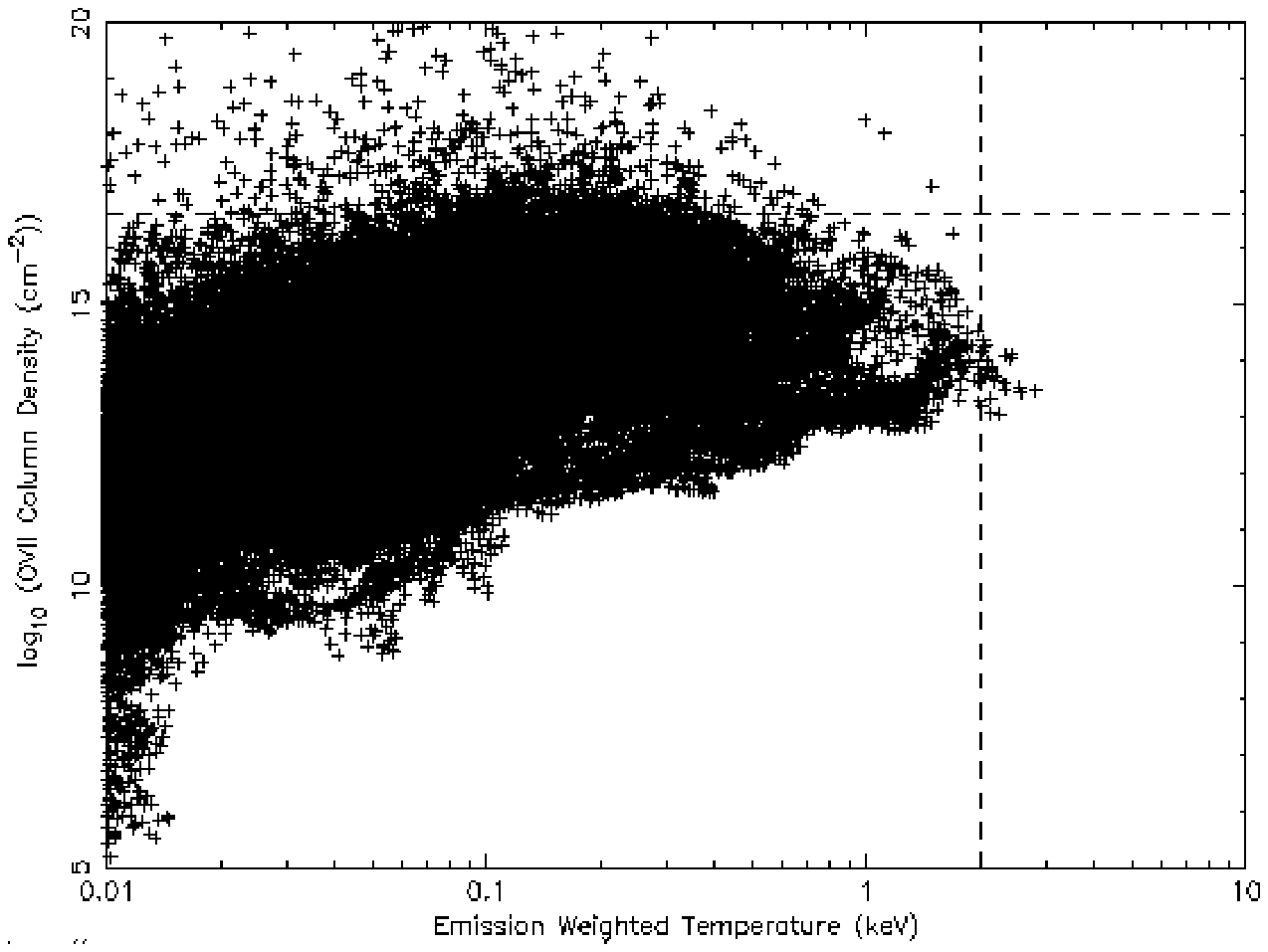}
\caption{The left panel and right panels show the distribution of \OVII\ column
  density as a function of the projected emission weighted temperature for both
  the small and large simulations respectively, In both plots the horizontal
  dotted line shows the observed lower limit to the \OVII\ column density
  derived from XMM-Newton observations of the CSE.  The vertical line shows the
  2 keV point above which we have assumed that that particular location will
  correspond to a cluster.  A CSE observation will correspond to point to the
  right and above the two lines but in both models there are far fewer cases
  (0\% and 1.26\% for the small and large model respectively) than would
  satisfy the observations}\label{ovii}
\end{figure}

The first possibility, that of large scale soft emission, has already been
demonstrated as being inconsistent with the models since the density of the
WHIM and hence the emissivity is too low.  We investigated the second argument
using both the small and large scale simulations of Cen and Ostriker (scale 25
and 100 Mpc respectively) and have calculated the \OVII\ column densities
projected along the line of sight.  In both cases we then compared the emission
weighted temperature with the \OVII\ column density to see where and how often
the \OVII\ is consistent with the strength of the line emission reported by
Kaastra et al.  The two models are shown in figure~\ref{ovii}, and indicate
that at locations where the projected emission weighted temperature is greater
than 2 keV (which corresponds to locations where a cluster would exist) very
few cells have the required \OVII\ column.  Indeed, for the smaller of the two
models (the model with the higher resolution) there were no areas with an
\OVII\ column $> 4 \times 10^{16}$ cm$^{-2}$ (the density required by Kaastra
et al. 2003) and a temperature above 2 keV.  For the larger, low resolution
model 45 cells out of a total of 3568 consistent (or 1.26\%) with a cluster had
an \OVII\ column large enough.  While this may indicate that there is some
relationship between a CSE the density of structures in the model, 1.26\% is a
very small fraction compared to the size and scale of the observed soft excess
evident in the real data.  From the point of view of absorption line studies,
these plots also show that in the vicinity of clusters we would expect an \OVI\
absorbing column of order $10^{14}$cm$^{-2}$ rather than the value predicted by
Kaastra et al. ($4 \times 10^{16}$cm$^{-2}$).  Therefore, based on these
simulations we would say that there is no strong correlation between emission
from the WHIM assoiciated with superclusters and the CSE.

\clearpage

\clearpage

\section{Other models}

All of the above discussion has been based on the analysis of one particular
model.  It is reasonable to ask if other models show the same effect.  This is
actually quite a difficult question to answer since there are only a few models
with all the required information freely available.  A larger volume, lower
resolution model with a size of 100 h$^{-1}$ Mpc on a uniform mesh with 512$^3$
cells with $256^3$ dark matter particles has also been analysed.  This model
has been derived from the same code as was used for the higher resolution model
discussed above.
In general we find the same result - it is very difficult to reproduce the
observed soft excess signal.  With a larger volume there is, of course, a
slightly higher probability of having galaxy groups located along any sight
line which can give rise to excess soft flux in a similar fashion to the case
studied in section~\ref{group}.  However, even by taking this effect into
account we still arrive at a similar conclusions to those obtained form the
higher resolution model - the WHIM itself cannot reproduce the observed CSE.
We have also studied one of the clusters available in the Laboratory of
Computational Astrophysics (LCA) simulated cluster archive
(http://sca.ncsa.uiuc.edu) with no major change in the conclusions.

\section{Conclusions}

From the above discussion it would appear that within our current understanding
there is insufficient luminosity and low temperature material ($< 1$\ keV) to
enable association of an observable CSE with a WHIM filament.  What are the
possible reasons for this discrepancy?  Perhaps the most obvious problem with
the simulations is the need for the highest resolution possible coupled with a
very large simulation volume.  A high resolution enables an accurate simulation
of the gas in small groups which may be important (see section~\ref{group}),
while a large volume will allow the simulation to accurately represent the WHIM
on the scale of filaments and superclusters, again an idea that has been
proposed to explain the CSE (Kaastra et al. 2003,2004 and
section~\ref{supercluster}).  Here we note that our high resolution simulation
is only accurate over scales of 5-6 Mpc which is smaller than the proposed size
of the Coma filament (Finoguenov et al. 2003), whereas with the low resolution
model one can investigate scales commensurate with a filament but not small
groups.  Therefore the current model may very well have underestimated the soft
X-ray/CSE emission from the WHIM.  It is, however, inevitable that any
difference between current and future models will be small when compared to the
size of the CSE signal, since the current discrepancy between models and
observation is so large (10-30 times by mass).  This is certainly too large a
difference to be a resolution or scale issue.  Therefore, while new simulations
may be able to account for more of the CSE than current models, there is
currently no evidence that emission from the WHIM is the cause of the bulk of
the CSE signal.

Given it is unlikely for the WHIM as modeled to be associated with the CSE
signal, what other options are we left with?  If we assume a thermal
intracluster origin of the CSE we are back to the case where we have
significant problems with gas cooling times, since the density of the warm ICM
must then be high enough for it to be in pressure equilibrium with the hot ICM.
As mentioned in the introduction, this would then require an efficient heating
mechanism to replenish the radiated energy, which is not easy to envisage.  On
the other hand, in the context a non-thermal interpretation of the CSE the
XMM-Newton detection of OVII emission lines in the CSE spectrum remains
unexplained (unless, of course, the observational reality of such lines is
called to question).  Another formidable challenge to the non-thermal model
concerns the \ROSAT Coma results of Bonamente et al. (2003), where the CSE is
reported to exist on a very large spatial scale.  Thus the true nature of the
soft X-ray emission component in clusters of remains illusive.

\acknowledgments
JPDM and RL acknowledge the support of NASA grants NAGS-13095 and NAGS-11484.

\end{document}